\documentclass{pasj01}
\draft
\Received{$\langle$reception date$\rangle$}
\Accepted{$\langle$acception date$\rangle$}
\Published{$\langle$publication date$\rangle$}

\begin{document}

\title{Time resolved spectroscopic observations of an M-dwarf flare star EV~Lacertae during a flare}
\author{Satoshi Honda\altaffilmark{1}, Yuta Notsu\altaffilmark{2}, Kosuke Namekata\altaffilmark{2}, Shota Notsu\altaffilmark{2}, Hiroyuki Maehara\altaffilmark{3},  Kai Ikuta\altaffilmark{2}, Daisaku Nogami\altaffilmark{2}, and Kazunari Shibata\altaffilmark{2,4}}

\altaffiltext{1}{Nishi-Harima Astronomical Observatory, Center for Astronomy, University of Hyogo, 407-2, Nishigaichi, Sayo-cho, Sayo, Hyogo 679-5313, Japan}
\altaffiltext{2}{Department of Astronomy, Kyoto University, Kitashirakawa-Oiwake-cho,Sakyo-ku, Kyoto 606-8502}
\altaffiltext{3}{Okayama Observatory, Kyoto University, 3037-5 Honjo, Kamogata, Asakuchi, Okayama 719-0232, Japan}
\altaffiltext{4}{Astronomical Observatory, Kyoto University, Kitashirakawa-Oiwake-cho, Sakyo-ku, Kyoto
606-8502, Japan}
\email{honda@nhao.jp}


\maketitle

\begin{abstract}

We have performed 5 night spectroscopic observation of the H$\alpha$ line of EV Lac with a medium wavelength resolution ($R
\sim$ 10,000) using the 2m Nayuta telescope at the Nishi-Harima Astronomical Observatory.
EV Lac always possesses the H$\alpha$ emission line; however, its intensity was stronger on August 15, 2015 than during other four-night periods.
On this night, we observed a rapid rise ($\sim$ 20min) and a subsequent slow decrease ($\sim$ 1.5h) of the emission-line intensity of H$\alpha$, which was probably caused by a flare.
We also found an asymmetrical change in the H$\alpha$ line on the same night.
The enhancement has been observed in the blue wing of the H$\alpha$ line during each phase of this flare (from the flare start to the flare end), and absorption components were present in its red wing during the early and later phases of the flare.
Such blue enhancement (blue asymmetry) of the H$\alpha$ line is sometimes seen during solar flares, but only during the early phases. Even for solar flares, little is known about the origin of the blue asymmetry.
Compared with solar-flare models, the presented results can lead to the understanding of the dynamics of stellar flares.

\end{abstract}

\section{Introduction}\label{sec:intro}
Solar flares are energetic explosions in the solar atmosphere around active regions.
They are considered to be caused by the energy release through magnetic reconnections in the corona (e.g., \cite{Shibata2011}).
The flare energy, which is released in the corona and transported into the lower atmospheres, is dissipating in the dense chromosphere. This produces intense heating, which in turn leads to the evaporation and condensation processes.
Flares are observed across the entire electromagnetic spectrum from radio emission to high-energy gamma-rays.
Numerous observations of solar flares have been carried out in the coronal (e.g., X-ray) and chromospheric emission (e.g., H$\alpha$) bands so far (e.g., Shibata \& Magara 2011).
However, many of their aspects are yet to be investigated, both observationally and theoretically.

It is known that stellar flares occur on various types of stars.
They are observed mainly as a rapid rise of the emission intensity (X-ray, optical, radio, etc.) followed by its slow decrease.
In particular, young stars, close binary stars, and dMe stars can produce frequent flares (e.g., \cite{HawleyPettersen1991}, \cite{Kowalski2010}), and sometimes produce ``superflares'', i.e., flares whose total energy reaches values 10-10$^{6}$ times larger than the largest flares observed from the Sun.
Recent photometric data from the Kepler spacecraft found many superflare events on G,~K,~M-type stars (cf. \cite{Maehara2012}, \cite{Shibayama2013}, \cite{YNotsu2013}, \cite{Candelaresi2014}, \cite{Hawley2014}, \cite{Davenport2016}, 
\cite{Yang2017}).
These studies show many statistical properties of superflares.
The energy-release process of such large stellar flares can be explained by magnetic reconnection (\cite{Shibata2002}; \cite{Maehara2015}; \cite{Namekata2017}).
However, several issues remain to be investigated, especially concerning the dynamics and radiation mechanisms present during such large flares.

Numerous spectroscopic studies of solar flares have been performed in order to understand the flare dynamics and their radiation mechanisms.
The H$\alpha$ line is the most frequently observed line among solar flare spectroscopic observations.
Red asymmetry (enhancement of the red wing) has been often observed in the H$\alpha$ line profile during solar flares (e.g., \cite{IchimotoKurokawa1984}, \cite{Canfield1990}, \cite{Kuridze2015}).
This is thought to be due to the the process known as chromospheric condensation, which is the downflow of cool plasma in the flaring atmosphere.
Blue asymmetry has also been observed in the early phase of flares (e.g., \cite{Canfield1990}, \cite{Heinzel1994}).
This has been also observed in the Na~{\small I} D1, Ca~{\small II} 8542 \citep{Kuridze2016,Kuridze2017} and Mg~{\small II}~h line \citep{Tei2018}.
These results might mean the existence of cool plasma flows moving upward.
However, the detailed explanation of this blue asymmetry is still controversial. 

Such H$\alpha$ line asymmetries are also observed during stellar flares.
\citet{Houdebine1993} found red asymmetries in the core and wings of Balmer lines during a flare from a late-type dMe star (AD Leo).
They interpreted this as an evidence of downward chromospheric condensations, similar to those seen on the Sun.
An example of asymmetries present in the blue wing was found 
during a flare from the dMe star AT Mic in Balmer and Ca~{\small II} H and K line
 \citep{Gunn1994b}.
In addition, blue asymmetry has been sometimes observed during stellar flares (e.g., YZ CMi; \cite{Gunn1994a}, AD Leo; \cite{Crespo-Chacon2006}, and DG CVn; \cite{Caballero-Garcia2015} ).
With regards to YZ CMi, blue asymmetry has been observed also the in Mg~{\small II}~h line \citep{Hawley2007}.

However, previous spectroscopic observations of stellar flares are not sufficient to understand the mechanisms responsible for the red and blue asymmetries.
We need detailed spectra with higher temporal resolution, which has not been shown in previous studies. 
This is because flares themselves are rare events and high-temporal resolution observations require continuous observation during a time-span longer than a few hours.

EV Lac (GJ 873) is a well-known M4.5eV single-flare star.
From this star, superflares were also detected by X-ray (ASCA ; \cite{Favata2000}) and Gamma-ray (Swift ; \cite{Osten2010}) observations.
\citet{Schmidt2012} showed the frequency of flares from EV Lac to be 0.094 (events/hour), based on photometric observations.
These facts indicate that EV Lac is a promising target for the investigation of stellar flares.

In this paper, we present the results of spectroscopic observations of flares from EV Lac.
The details of the spectroscopic observations are described in Sec.2.
In Sec.3, we show the change of H$\alpha$-equivalent widths and line profiles with high temporal resolution.
These changes of the H$\alpha$ line are discussed in Sec.4.

\section{Spectroscopic observations of EV Lac}\label{sec:tarobs}

\begin{figure}[h]
  \includegraphics[width=8cm,angle=-90]{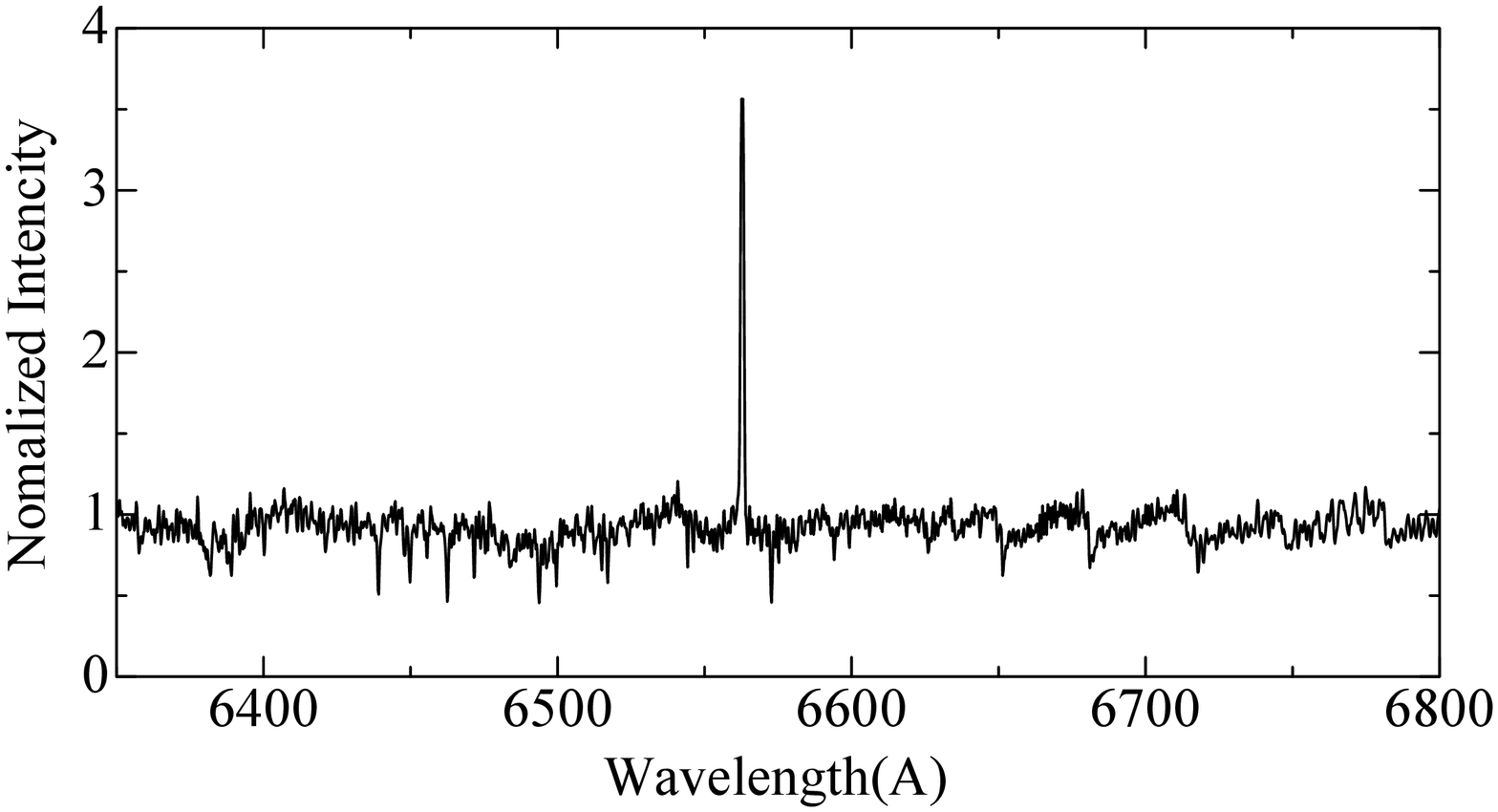}
\\
 \caption{Quiescent spectrum of EV Lac taken by MALLS on July 31, 2015. The strong emission line at the center of the spectrum is H$\alpha$. There are many atomic absorption lines and molecular bands present in this region.}\label{fig:spectra1}
\end{figure}

The spectroscopic observations were carried out by the Nayuta 2m telescope at the Nishi-Harima Astronomical Observatory.
The MALLS (Medium And Low-dispersion Long-slit Spectrograph) was used with a resolving power ($R=\lambda/\Delta\lambda$) of $\sim$ 10,000 at 6500{\AA}.
The wavelength range was 6350--6800~{\AA}.
This region includes the H$\alpha$ (6562.8~{\AA}) and He~{\small I} (6678.2~{\AA}) 
lines.
The observation runs were carried out on July 31, August 1, 15, 26, and 27, 2015.
The integration times for the single spectra (per frame) were 3 or 5 minutes.
The signal-to-noise ratio ({\it S/N}) of each observed spectrum was more than 50.
The total duration of the observations were 1, 2.5, 5.5, and 3.5 hours on August 1, 15, 26, and 27, respectively.
On July 31, only one frame of observation was conducted.
The log of observations is shown in Table \ref{table:obslog}.
Data reduction was done using the twodspec package of the 
IRAF\footnote{IRAF is distributed by the National Optical Astronomy Observatories,
which is operated by the Association of Universities for Research in Astronomy, Inc., under cooperate agreement with the National Science Foundation.} software in the standard manner (overscan correction, flat fielding, aperture determination, spectral extraction, sky subtraction, wavelength calibration, and normalization by the continuum).
Figure \ref{fig:spectra1} shows the spectrum of EV Lac obtained by MALLS on July 31 (quiescent phase, as described in Section 3.1).
We can see the strong H$\alpha$ emission line, and many atomic lines, molecular bands (TiO, CaH, etc.).
All observed spectra show similar features.

\section{Measurements of H$\alpha$ emission line}
\subsection{Equivalent width of H$\alpha$ emission line}
We measured the equivalent width (E.W.) of the H$\alpha$ line in every spectrum in order to investigate the flare and stellar activity.
For the measurement of the equivalent widths, the ``e" command of the splot task in IRAF was used.
We assumed the range of the H$\alpha$ emission line is from 6557.5~{\AA} to 6567.5~{\AA}, and integrated the flux above the continuum to measure the equivalent widths.
The measured values of the eqivalent widths are shown in Table 2.

\begin{figure}[htbp]
  \includegraphics[width=8cm,angle=-90]{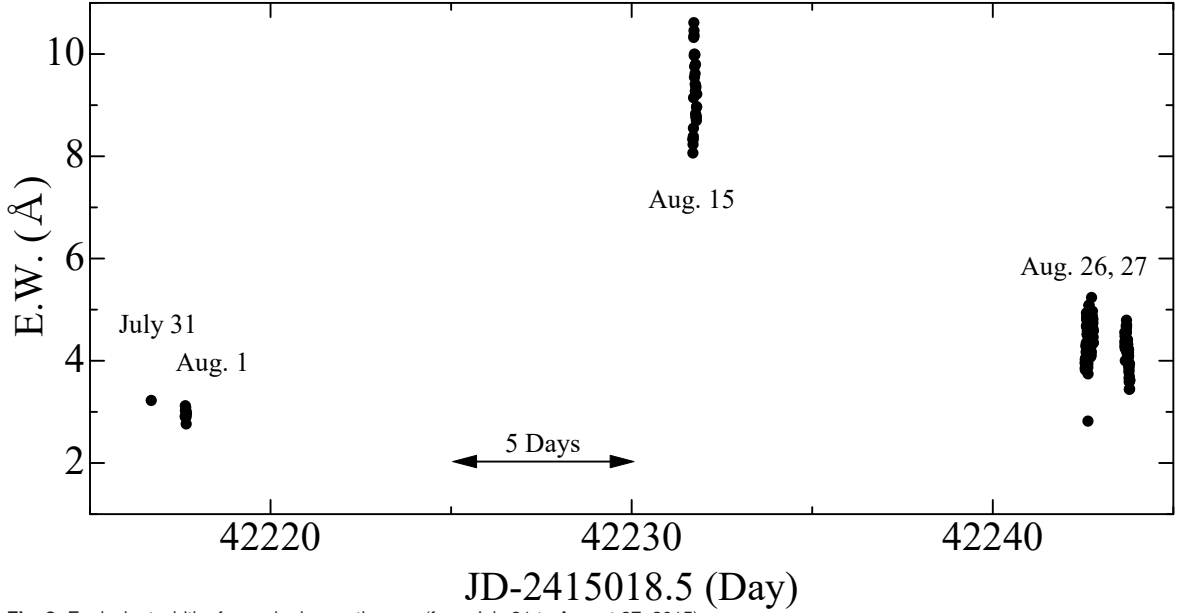}
   \caption{Equivalent widths for each observation run (from July 31 to August 27, 2015).}\label{fig:EW1}
\end{figure}

\begin{figure}[htbp]
  \includegraphics[width=5cm,angle=-90]{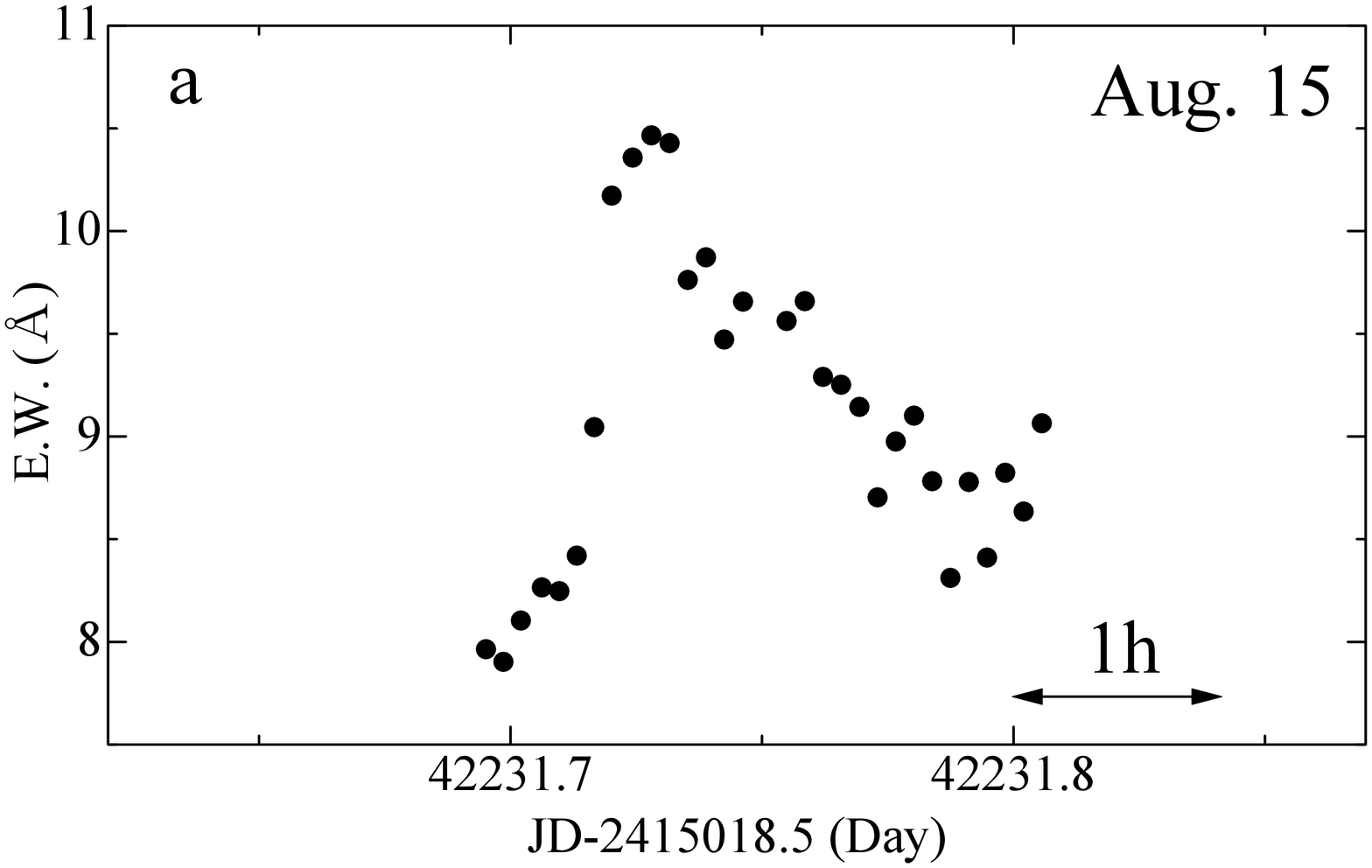}
  \includegraphics[width=5cm,angle=-90]{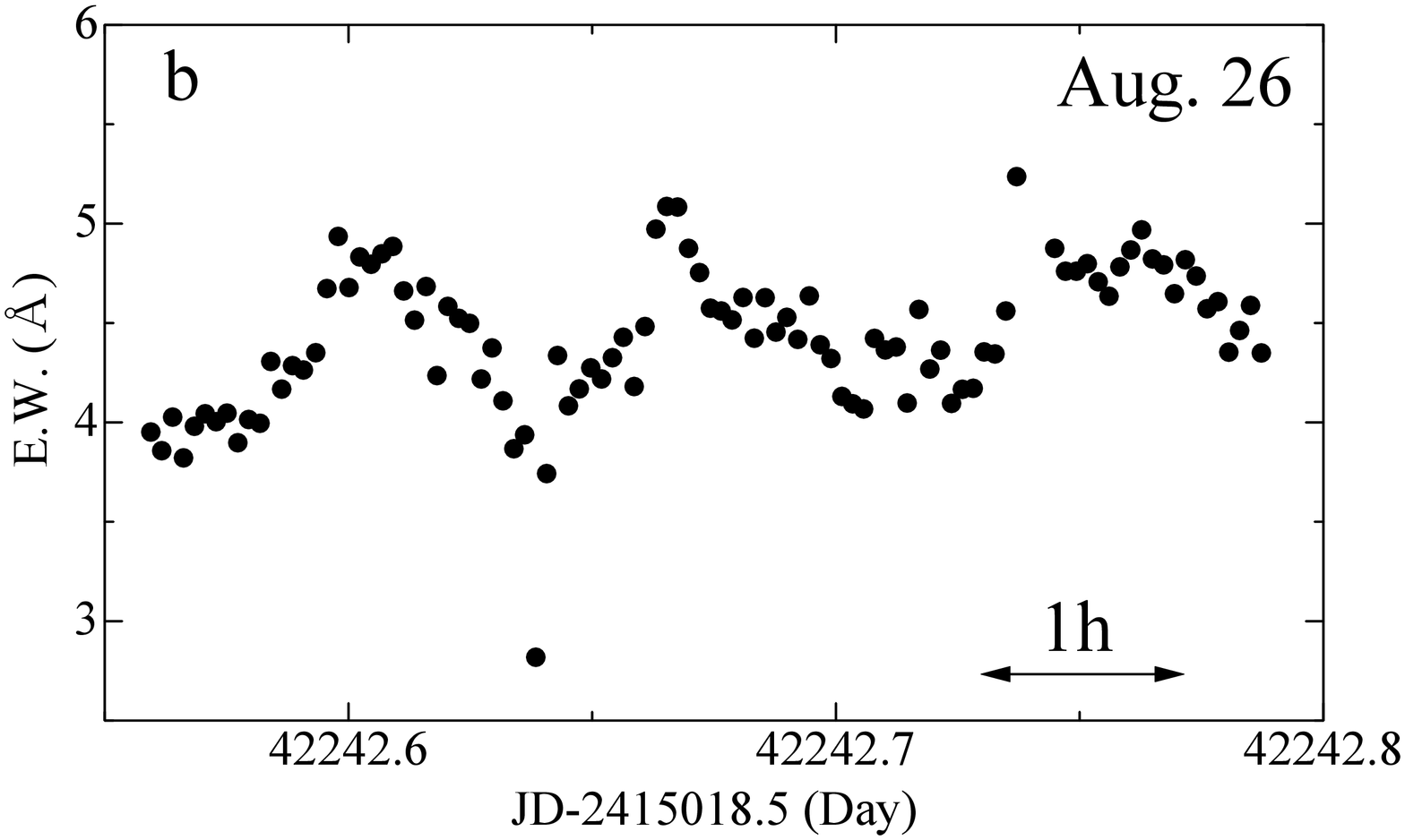}
  \includegraphics[width=5cm,angle=-90]{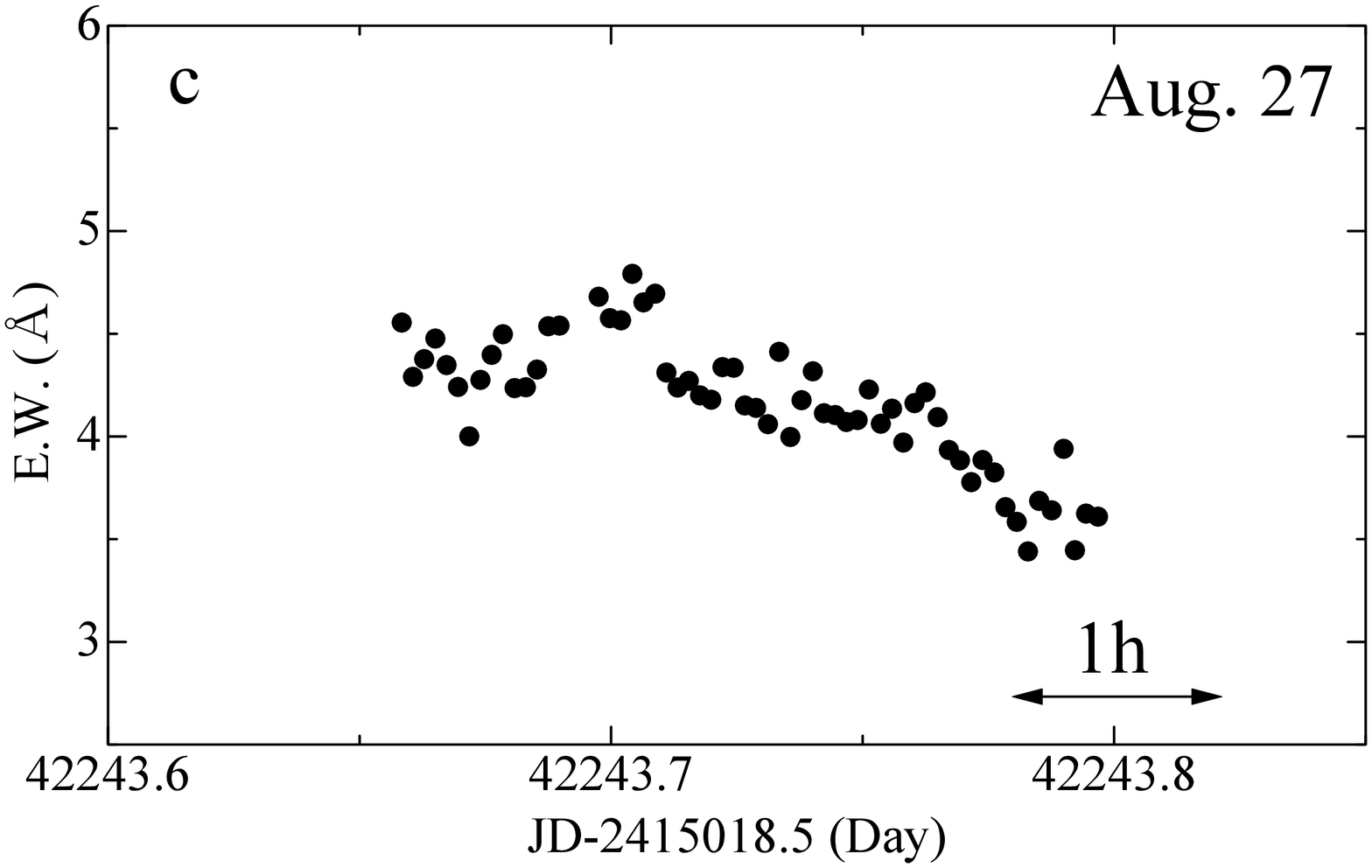}
   \caption{Equivalent widths for each observation run (August 15, August 26, and August 27 respectively). August 26 observation (b) shows a short-term modulation. This period is much shorter than the stellar rotation period (4.4 days). }\label{fig:EW2}
\end{figure}

In Figures \ref{fig:EW1} and \ref{fig:EW2}, we show the equivalent width of the H$\alpha$ line as a function of time for observation run.
In the July 31 and August 1 spectra, the equivalent width did not show significant changes.
The typical value of the equivalent width was about 3 {\AA} during these two periods.
\citet{Baranovskii2001} indicated that the H$\alpha$ equivalent widths of EV Lac are 3.5--5.3 {\AA} in the quiescent state and 4.7--10.6 {\AA} in the active state.
On August 1, we have obtained the 10 frames continuously during a 1-hour period, with no observable changes of H$\alpha$ intensity.
EV Lac was in a quiescent state during this period.

\begin{figure}[htbp]
  \includegraphics[width=9cm,angle=-90]{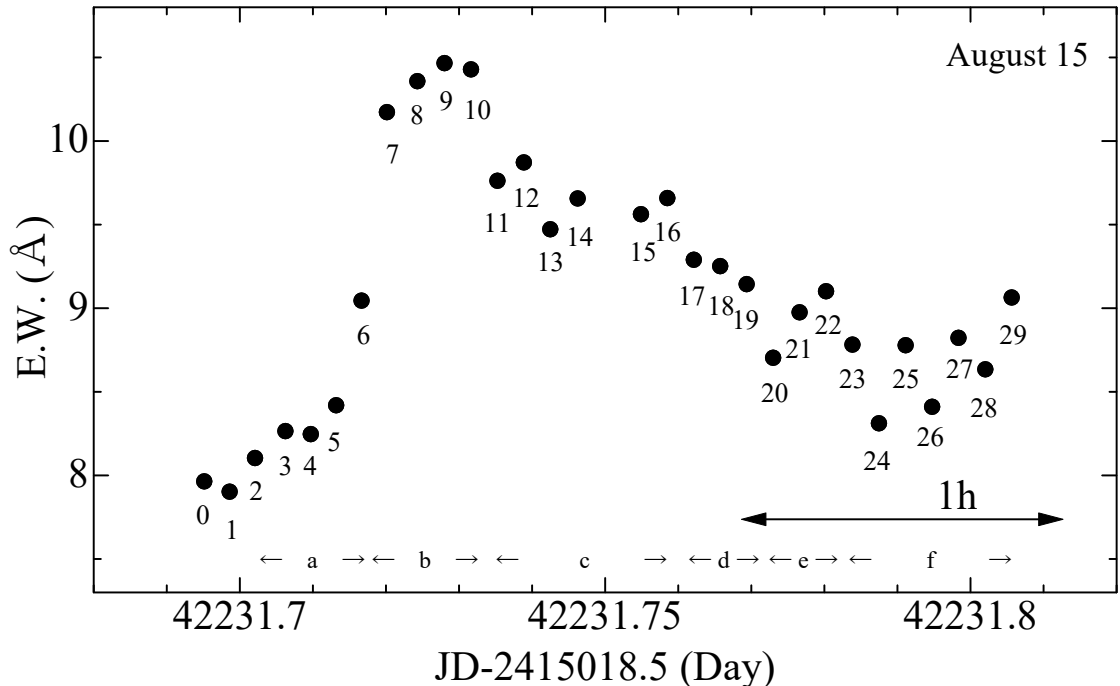}
   \caption{Same as August 15 of figure \ref{fig:EW1}, with each measurement labeled by numbers and the groups (see text for the details).}\label{fig:EW815num}
\end{figure}

\begin{figure}
  \includegraphics[width=8cm,angle=-90]{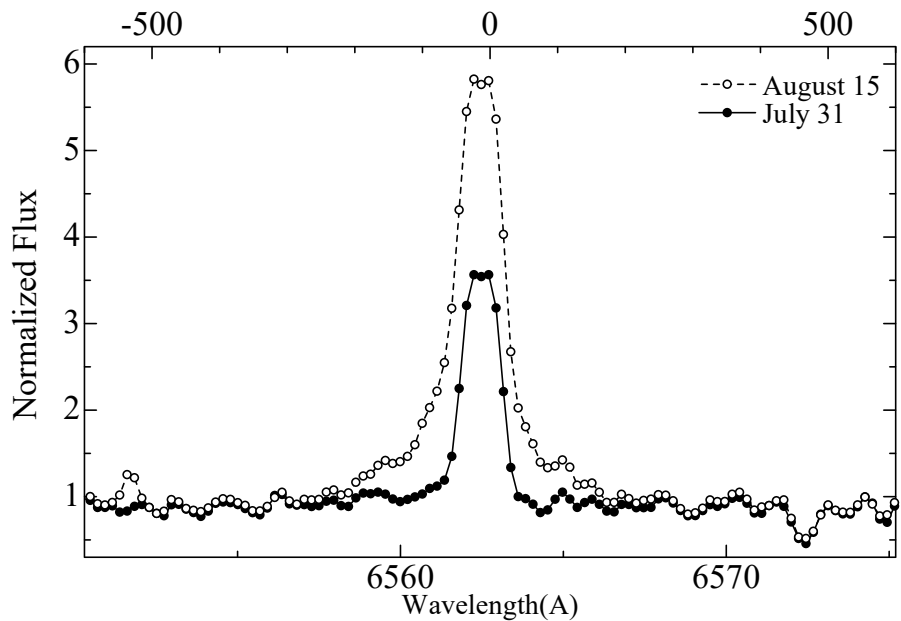}
\\
 \caption{Comparison of two spectra, from July 31 and August 15 (No.9 in Figure \ref{fig:EW815num}), respectively.
 The velocity is set to 0 km s$^{-1}$ at the H$\alpha$ center (6562.81{\AA}).}\label{fig:comp731-815}
\end{figure}

\begin{figure}[htbp]
  \includegraphics[width=9cm,angle=-90]{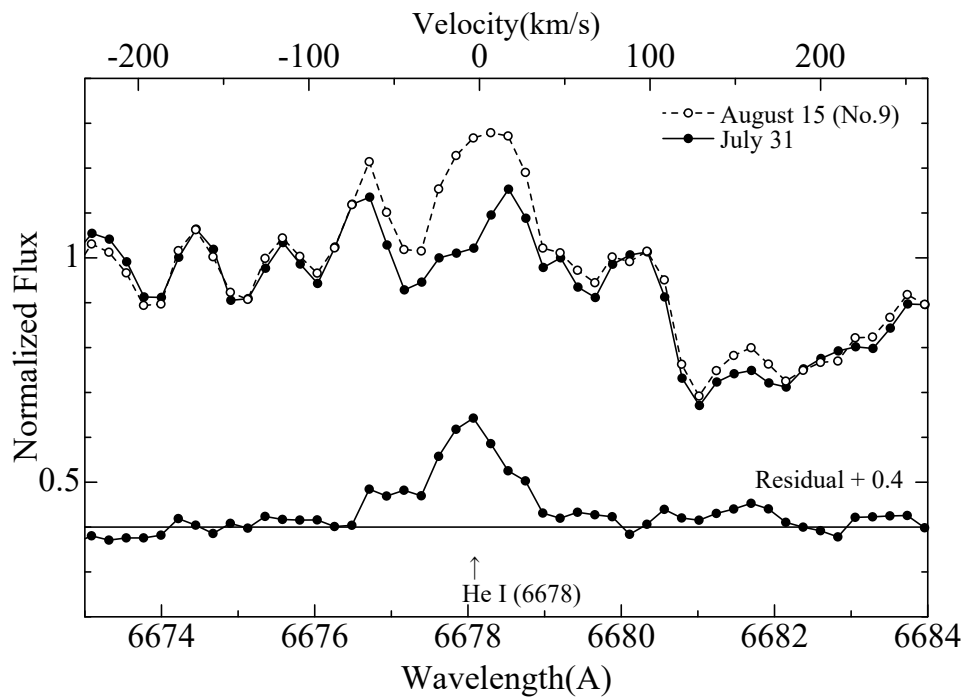}
\\
   \caption{Spectral comparison in the region around He I 6678.15 {\AA} and the residual spectrum obtained by subtracting the July 31 spectra from August 15 (No.9). The value of the vertical axis of the residual spectrum is shifted by +0.4. The velocity is set to 0 km s$^{-1}$ at the He I line center (6678.15{\AA}).}\label{fig:He}
\end{figure}

However, we found a substantial change of intensity of the H$\alpha$ line between July 31 and August 15.
Figure \ref{fig:comp731-815} shows the comparison of two spectra, July 31 and August 15 (No.9 in Figure \ref{fig:EW815num}), in the region of the H$\alpha$ line.
In the August 15 spectra, the equivalent width of the H$\alpha$ line showed large variations in the range of 8.1--10.5 {\AA}.
These results suggest that EV Lac was in an active state that night.
In addition, the H$\alpha$ line showed a rapid rise followed by a slow decrease of intensity, which was probably caused by a long-duration flare.
The enhanced H$\alpha$ emission is a common feature of solar and stellar flares (e.g., \cite{Haisch1991}).
It increased by 2 {\AA} within 20 minutes and then slowly decreased during a period of 1.5 hours.
This is a similar shape as possessed by typical light curves of solar flares.
However, the present observation was a very long duration flare.
The He {\small I} 6678 {\AA} line is also present in each spectrum taken this night, although it was not observed on other nights.
Figure \ref{fig:He} shows the spectra of the region around He {\small I} 6678 {\AA}.
This He emission can be a further evidence of the flare occurrence.
The He {\small I} 5876 {\AA} D3 line is well-known to show emission during flares.
The He {\small I} 6678 {\AA} line is expected to show the same behavior as the D3 line (e.g., Eason et al. 1992).
However, it was too weak to discuss its changes of intensity during this night.
We could not find any changes in other atomic and molecular lines.

The equivalent widths of H$\alpha$ line of August 26 and 27 were smaller than those of August 15.
However, small changes were seen on those days as well (Figure \ref{fig:EW2}).
From the August 26 observation, it seems that three small flares occurred.
However, no flares were seen during the August 27 observation.
A typical value of the equivalent width of the H$\alpha$ line was 4.5 {\AA} on those two days, which is in the range of the ``quiescent state"  \citep{Baranovskii2001}.
Although the equivalent width of H$\alpha$ shows a gradual decrease on August 27, it might be due to the change of appearance of an active region with the rotation of a star.
However, we also note that \citet{Pettersen1980} suggest that the variations of the H$\alpha$ line do not correlate clearly with the rotation.

\subsection{Profiles of H$\alpha$ line during August 15 flare}
\begin{figure}[htbp]
  \includegraphics[width=9cm,angle=-90]{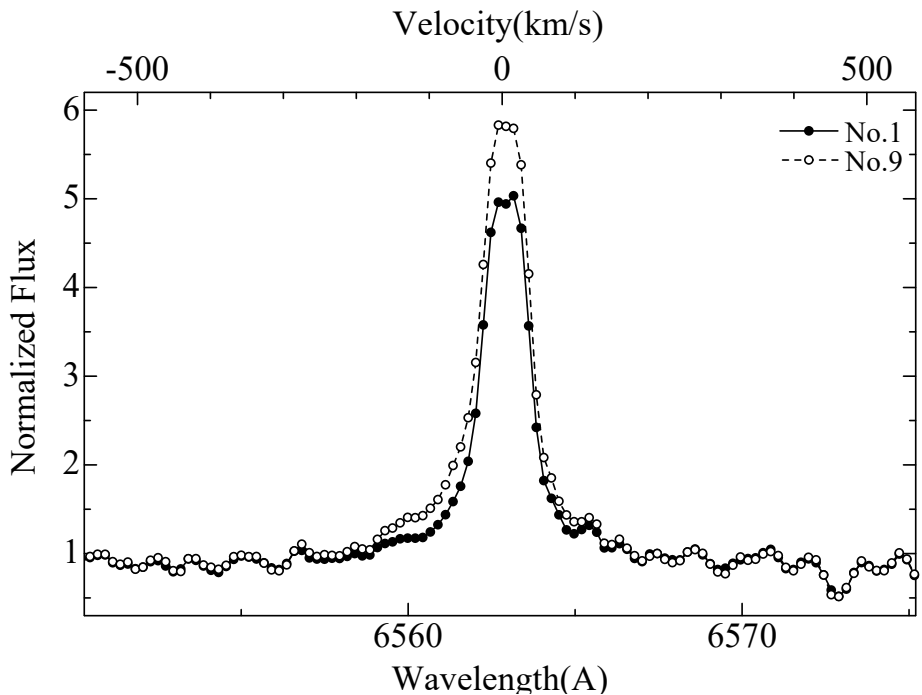}
  \\
 \caption{Comparisons of spectra taken on August 15. 
 The open circle and dashed line indicate the strongest line (No.9), the filled circle and solid line indicate the weakest line (No.1). 
 The time interval between No.1 and No.9 is 42 minutes.
 The upper axis shows the velocity, lower axis the wavelength. 
 The velocity is set to 0 km s$^{-1}$ at H$\alpha$ line center (6562.81{\AA}).}\label{fig:comp815_2-9}
\end{figure}

In the observations on August 15, we found a change in the H$\alpha$ line profile.
Figure \ref{fig:comp815_2-9} shows the comparison of the weakest and strongest features of the H$\alpha$ line on this night.
The time interval between No.1 and No.9 is 42 minutes.
No clear shift of the line center wavelength was found.
We found an enhancement of the wing component of the H$\alpha$ line in the blue region. The flux enhancement in the blue region around $-$100 km s$^{-1}$ is clearly lager than that in the red region around $+$100 km s$^{-1}$ (Figure 7, see also Figure 9b).
The change in the line profile during the flare was stronger in the blue wing component of the line than in the red one. 
We discuss about this in detail in Section 4.

Previous studies \citep{Johns-Krull1997} observed a solar flare with high-resolution spectroscopy and found that the H$\alpha$ emission line during the flare showed a double-peak profile.
\citet{Martin1999} showed that the H$\alpha$ line of the dMe star VB~8 has a double peaked profile, showing the variation of peak intensity ratio.
Before the flare, the blue-peak intensity is stronger than that of the red peak.
However, this peak intensity ratio is reversed during the flare.
After the flare, the blue peak becomes stronger than the red, once again.
Our observations of EV Lac also show the features of a double-peak profile for the H$\alpha$ line (Figure \ref{fig:comp815_2-9}).
The spectrum from July 31 shows the features of the same peak intensity ratio.
However, the line had a variation of double-peak profile with the peak of red part having a more prominent emission than the blue in an August 15 spectrum (No.1).
It seems that the peaks were reversed pre-flare and during the flare (Figure \ref{fig:comp815_2-9}).
This change of peak intensity ratio is the opposite of observations of VB~8 \citep{Martin1999}, but it is not clear because of the insufficient wavelength resolution of the spectra.
We must note that to compare the intensity ratio of double peaks, we need higher wavelength-resolution spectra.
In the observations by \citet{Martin1999}, the wavelength-resolution of the spectra ($R=$ 110,000) were higher than that in our observations ($R\sim$ 10,000).

\begin{figure}[htbp]
  \includegraphics[width=12cm]{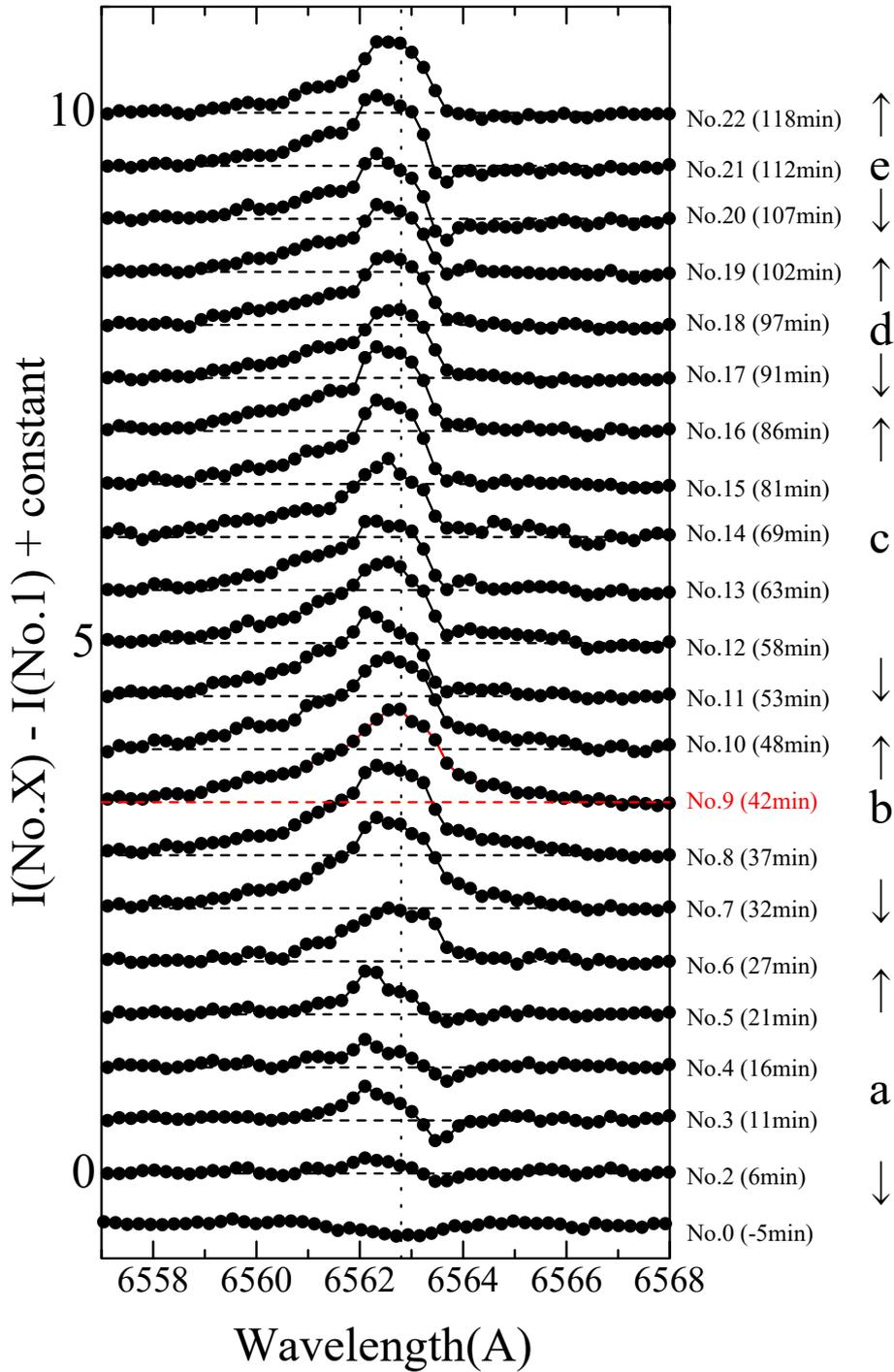}
   \caption{Time sequence of subtracted line profiles of the spectra taken on August 15. (No.0--22).}\label{fig:profiles2}
\end{figure}

\begin{figure}[htbp]
  \includegraphics[width=6cm,angle=-90]{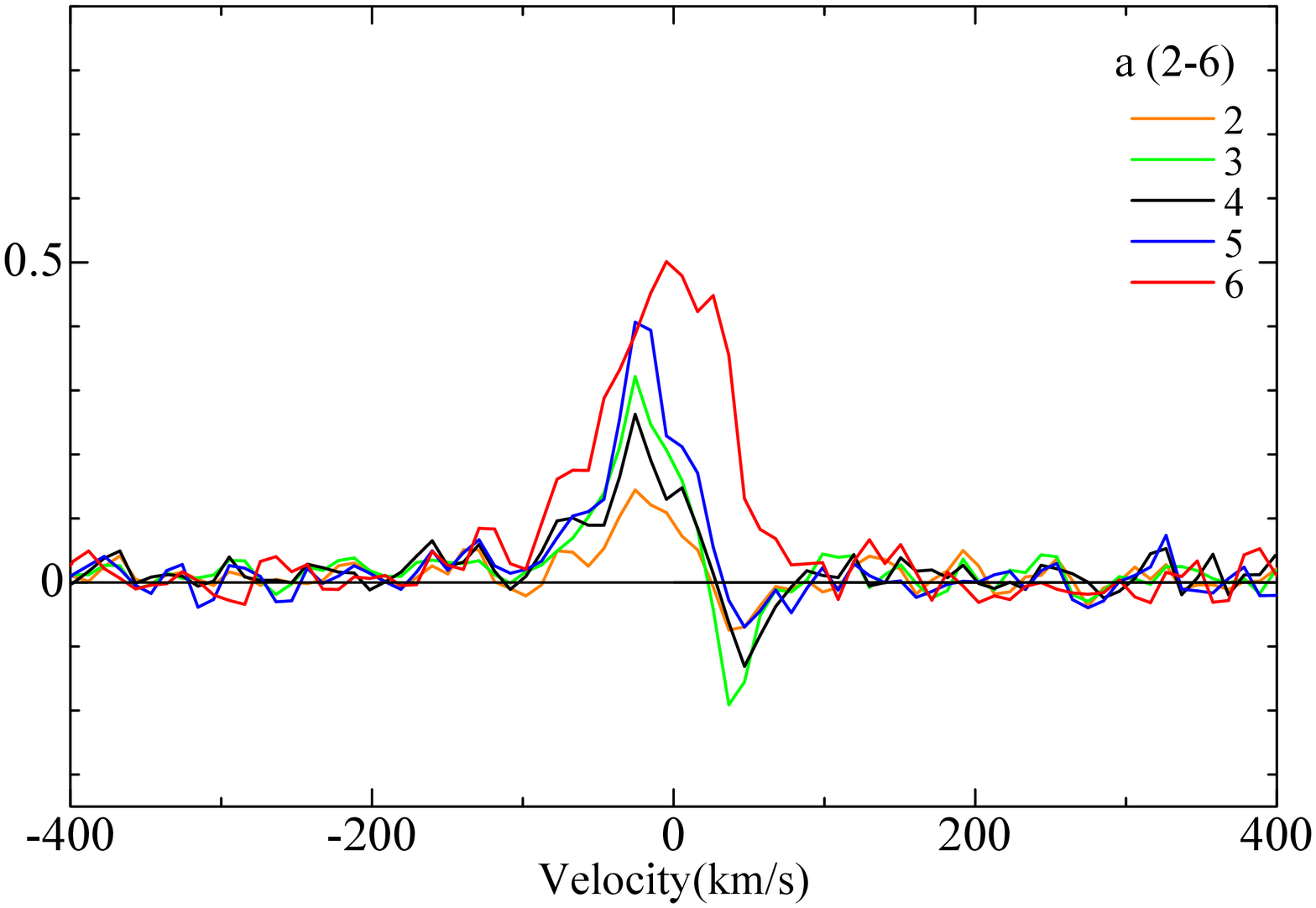}
  \includegraphics[width=6cm,angle=-90]{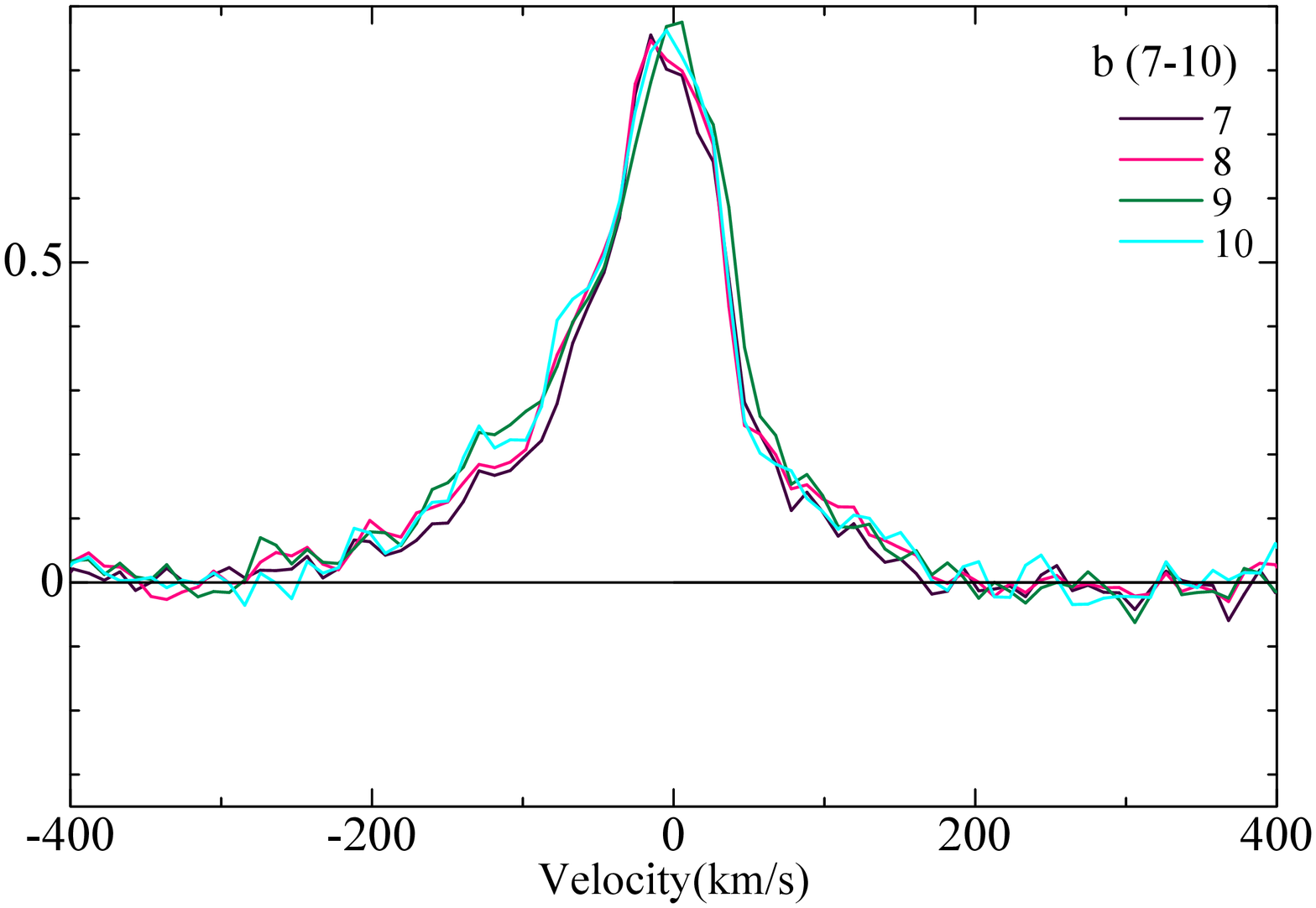}
  \\
  \\
  \includegraphics[width=6cm,angle=-90]{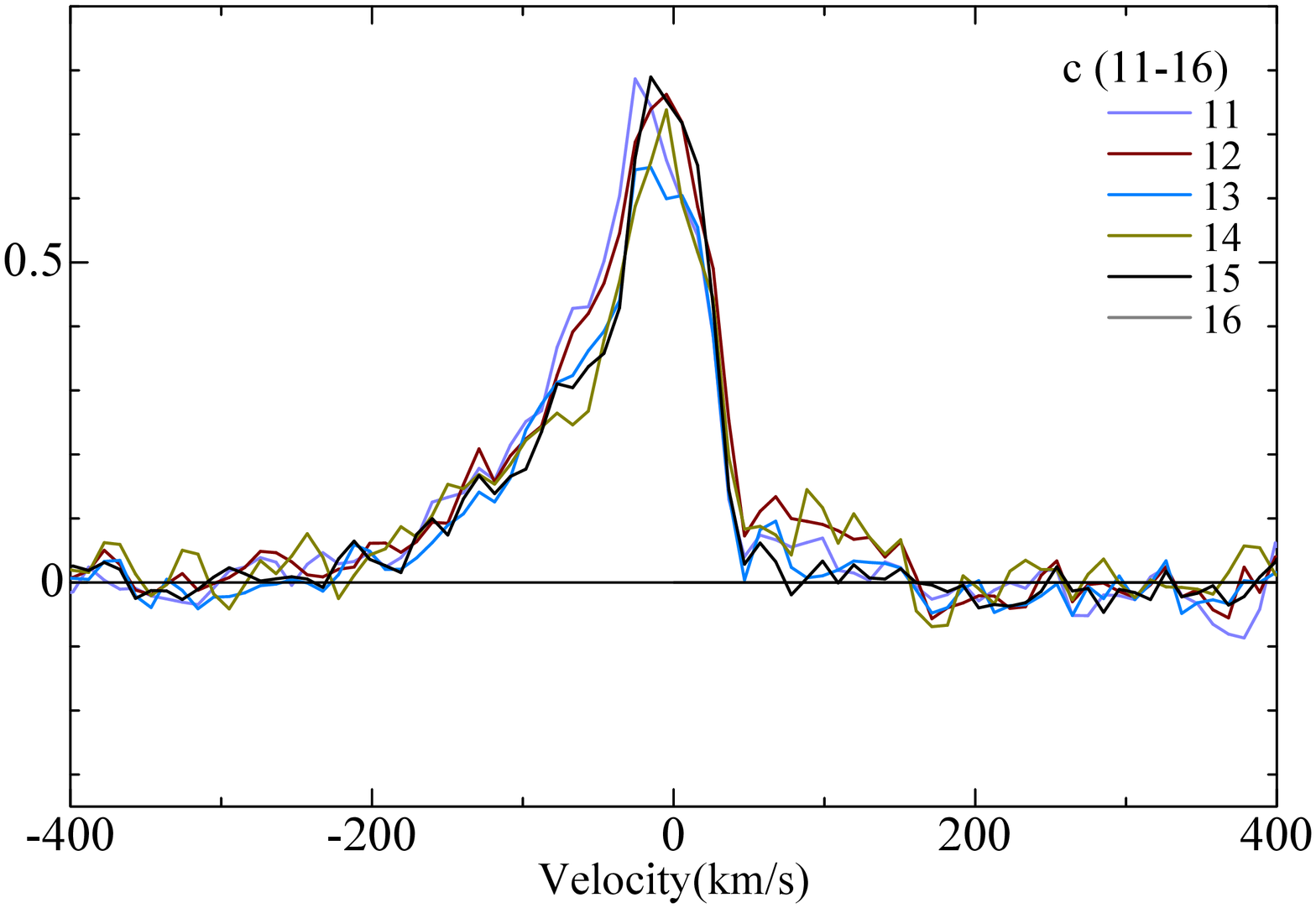}
  \includegraphics[width=6cm,angle=-90]{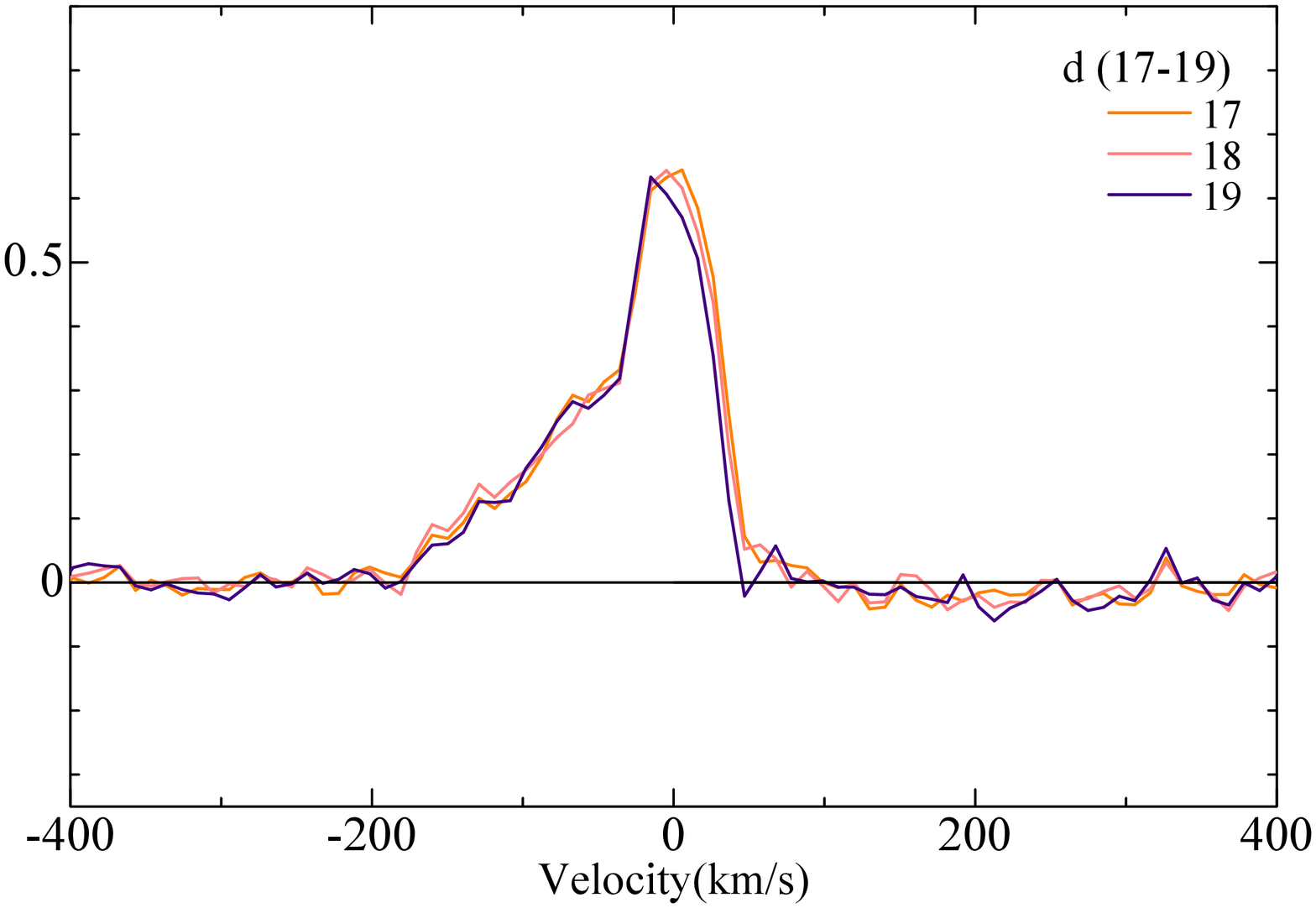}
  \\
  \\
  \includegraphics[width=6cm,angle=-90]{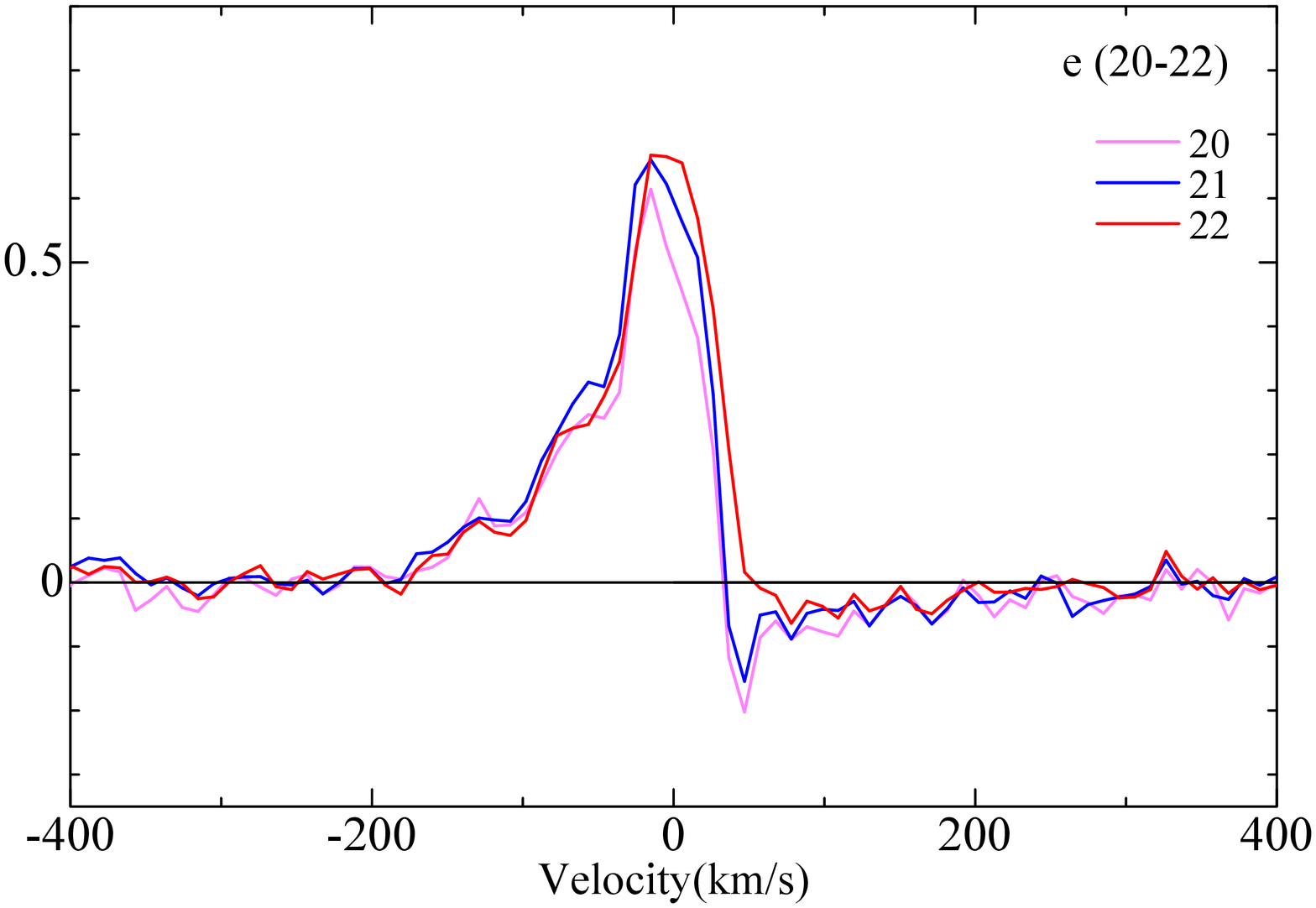}
  \includegraphics[width=6cm,angle=-90]{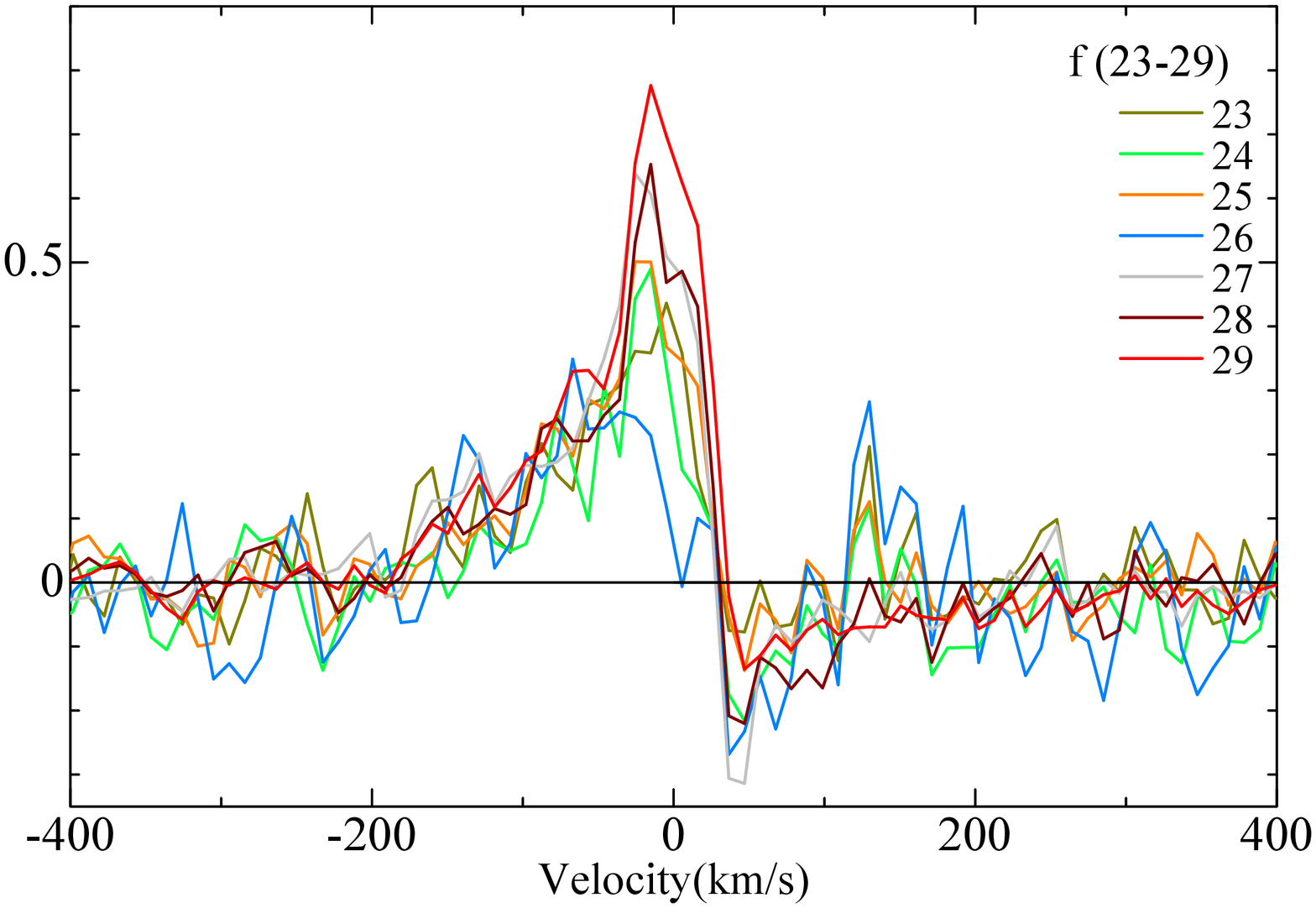}
     \caption{Same as figure \ref{fig:profiles2}, but the subtracted spectra are shown superimposed in each phase (cf., figure \ref{fig:EW815num}).
   a (2--6):impulsive phase, b (7--10):peak phase (E.W. $>$ 10 {\AA}), c (11--16): decay phase (10 {\AA} $>$ E.W. $\gtrsim$ 9.5 {\AA}), d (17--19):decay phase (9.5 {\AA} $>$ E.W.), e (20--22):small absorption in red part, f (23--29):poor quality spectrum.
    (Color version of these figures is available in the online journal.)}\label{fig:profiles-velocity}
\end{figure}

To clearly see the pure-flare components, we subtracted the pre-flare spectra.
We assumed the second spectrum (No.1 in Figure \ref{fig:EW815num}) of the series as a pre-flare spectrum, which showed the smallest equivalent width this night.
Figure \ref{fig:profiles2} 
shows the time series of the subtracted line profiles of H$\alpha$.
The subtracted spectra show the variations in the emission and absorption components.
Figure \ref{fig:profiles-velocity} shows that in the impulsive phase of the flare (Figure \ref{fig:profiles-velocity}a: No. 2--6 in Figure \ref{fig:EW815num}), an increase of the intensity was observed in the blue wing of the line, and an absorption component was observed in the red wing.
This absorption component with a velocity of a few tens of km s$^{-1}$ exhibited a change of intensity in time.
This absorption component does not correspond to the wavelength of the water vapor lines (6564.206 {\AA}).
In the 4th spectrum (No.3), 
the red absorption component exhibits its strongest features.
After that, the intensity of the blue component increased, together with the whole-emission component, while the red absorption decreased.
In the 7th spectrum (No.6), the absorption component disappeared.

In the peak-phase spectra (Figure \ref{fig:profiles-velocity}b), the profile did not change from the 8th (No.7) to 11th (No.10) spectra.
In those spectra, the values of the equivalent width are more than 10 {\AA}, and no absorption features are observed.

The 12th (No.11) or later spectra showed a deficit of the red wing and the variations of intensity in the line center (Figure \ref{fig:profiles-velocity}c : 10 {\AA} $>$ E.W. $\gtrsim$ 9.5 {\AA}, Figure \ref{fig:profiles-velocity}d : 9.5 {\AA} $>$ E.W.).

The absorption component in the red wing appeared again in the 21th (No. 20) and later spectra (Figure \ref{fig:profiles-velocity}e).
Although those spectra are noisy (Figure \ref{fig:profiles-velocity}f) because of bad weather, we can see that the enhancement of the blue part continued until the end of the observation.

\section{Discussion}\label{sec:discussion}
\subsection{Variabilities of equivalent width of H$\alpha$ emission}
The intensity of the H$\alpha$ line of EV Lac shows substantial changes.
We can consider the following three possibilities to explain the change of the H$\alpha$ intensity: 
(1) the brightening by the flares (typical time scale of $\sim$~ hours); (2) visibility of the active region with the rotation of the star (rotation period $=$ 4.4 days); (3) evolution of active regions (days $\sim$ months).
During our one-month observation period, EV Lac exhibited a change from its quiet phase to its active phase.
Since the rotation period of EV Lac is 4.4 days \citep{Pettersen1980}, roughly 50\%
 of the stellar surface visible to the observer on August 1 would also be visible on August 15.
This indicates that the change of the H$\alpha$ intensity might be caused by the visibility change of the active region via stellar rotation.
We could not determine the origin of the change of the H$\alpha$ intensity from our observations.
Several spectroscopic observations of EV Lac have been carried out with a focus on the H$\alpha$ line  (e.g., \cite{Alekseev2003}, \cite{Melikian2006}), but there are very few observations with time cadence similar to that of our observation.
If we aim to clarify the relationship between the fluctuations of line intensities and stellar activities, more frequent and long-term monitoring is necessary.

The large change in the equivalent width on August 15 may have been caused by a very long-duration flare.
The equivalent width has already become larger than previously observed by the beginning of the flare.
On this night, the active region of the stellar surface might have been large and flare might have occurred.
Otherwise, a few huge flares or a lot of small flares occurred on the stellar surface before the large flare we observed.
Simultaneous photometric data would be helpful to understand the variations of equivalent width of H$\alpha$.
Unfortunately, we do not have the data of photometric observations for the observed periods.
We note that we could not observe any filling-up of atomic lines even during the flare.

\subsection{Line profile of H$\alpha$ emission during the flare}
The change of the H$\alpha$ line profile is associated with the dynamics of the cool gas ($\sim$10,000 K) present in the flare.
Blue asymmetry (enhancement of the blue H$\alpha$ emission wing), which was seen in our observations, is also observed at the early phase of solar flares.
However, the mechanisms are still unknown.
Several studies have been made on line profile asymmetries are produced in the solar flare models.
Solar flare models obtained with the radiative hydrodynamic simulations show that the asymmetries in H$\alpha$ line could be produced by the strong velocity gradients in the chromosphere generated by the evaporation and condensation processes (e.g., \cite{AbbettHawley1999}, \cite{Kuridze2015}).
The velocity gradients create differences in the opacity between the red and blue wings of H$\alpha$, and the sign of the gradient determines whether the asymmetric emission appears to the blue or red side of the line profile.
One explanation of the blue asymmetry is that if the downflow occurs in the upper chromosphere, its effect would be a red wing absorption, which can lead to a blue asymmetry in the line profile (e.g., \cite{Canfield1990}, \cite{Gan1993}, \cite{Heinzel1994}, \cite{Kuridze2015}).
Another explanation is that a cool dense plasma moves upward via chromospheric evaporation and  filament activation during the early or late phase of the flare (e.g., \cite{Canfield1990}, \cite{Allred2006}, \cite{Huang2014}, \cite{Tei2018}).
These three models can explain solar flares whose blue asymmetry is seen only in the early or late phase.
However, the blue asymmetry in this observation is different from the solar one.
It was present during the whole period, from the early phase to the later phase of the flare.
The blue asymmetry in the decay phase can be explained by the absorption by downward-moving plasma of post-flare loops in the corona, although this model cannot explain the blue asymmetry in the early phase.
As we described above, there are many studies illustrating how blue asymmetry is generated during solar flares.
However, there is no definite proof of any of them, and possibly newly developed ideas are necessary for a more detailed understanding.
This observation can not determine the origin of the blue asymmetry, but it will give important clues for understanding flare dynamics and radiation mechanisms.

A sharp absorption component is seen in the red part of the H$\alpha$ line in the early and later phases of the flare during the observation.
The velocity of this absorption did not change during the flare, but the strength seemed to change.
This absorption component might be caused by plasma downflows produced (condensation patterns) in the post-flare loops.
In this case, the plasma flow produced by previous flares gets cold and descends.
\citet{Schmieder1987} shows the observed H$\alpha$ line profiles in post-flare loops on the Sun, 
and similar absorption features can be seen with similar velocity components as we observed.
\citet{Canfield1990} also shows the blue- and red-shifted absorption features in an eruption of a  filament on the Sun in the early phase of a flare.
\citet{Cao2012} found several absorption features of the H$\alpha$ region in the subtracted spectra of a very active RS CVn-type star.
The absorption features can be explained by cool post-flare loops projected against the bright flare background, which sometimes occur during the gradual decay phase of large two-ribbon solar flares and can last for up to several hours \citep{Wiik1996,Jing2016}.

We discussed the observed stellar flare considering several solar flare studies.
On the observations of solar flares, it is performed in various wavelength regions with high spatial resolutions, and detailed information are obtained.
In order to apply the solar flare models to the stellar flare, more detailed observations are required.
The detailed observation with time and spatial cadence comparable to the solar case is difficult for the observations of stellar flares, but high-dispersion spectroscopy and multi-band observations will be useful for the studies of flare mechanism in detail.

\section{Summary}
We observed the dMe flare star EV Lac with a medium-resolution spectrograph mounted on the Nayuta telescope.
The equivalent width of the H$\alpha$ emission showed substantial changes during our observation period.
On August 15, 2015, we observed a rapid rise and the following slow decrease of the emission-line intensity of H$\alpha$, which was probably caused by a flare.
The profiles of H$\alpha$ during the flare exhibited blue asymmetry and an absorption component in the  red wing of the emission line.
In contrast to the solar flares where blue asymmetries are mainly reported during the early phase of the flare, we observed it during the whole duration of the flare. 
We note that the blue asymmetry has also been produced during the late stage of solar flare events due to the filament activation/eruption and evaporation.
We speculate that the blue asymmetries in EV Lac presented in this work is likely to be the analog to solar observations that report the blue asymmetries in the flare profiles.
The absorption component seen in the red part of the emission may be caused by plasma downflows produced (condensation patterns) in the post-flare loops.
However, there is no definite proof of either interpretation at the present.
In order to understand the gas dynamics during flares, spectroscopic observations with higher spectral and temporal resolution are necessary.

\bigskip
\bigskip
\begin{ack}

We would like to thank Dr. Petr Heinzel for the discussions on the interpretation of the results, Drs. Suzanne L. Hawley, and Adam F. Kowalski also, for their helpful comments on the work.
We also thank the staff of NHAO.
We also thank an anonymous referee for many helpful comments.
This work was supported by the Grant-in-Aids from the Ministry of Education, 
Culture, Sports, Science and Technology of Japan (No. 26400231, 16H03955, 16J00320, 16J06887, 17K05400, and 17H02865).
\end{ack}

\bigskip
\bigskip

\begin{table}[htbp]
  \caption{Observing Log}\label{table:obslog}
    \begin{tabular}{llcccc}
      \hline
      \hline
Obs. date & Start-End time(JST) & Observation Duration (hours) & Exp. Time (sec) & Number of spectra &  \\
      \hline 
2015-07-31 & 25h 46m  - 25h 56m & 0.17h & 600s &  1  &  \\
2015-08-01 & 27h 22m  - 28h 15m & 1h    & 300s & 10  &  \\
2015-08-15 & 25h 41m  - 28h 26m & 2.5h  & 300s & 30  &  \\
2015-08-26 & 22h 25m  - 27h 56m & 5.5h  & 180s & 100 &  \\
2015-08-27 & 24h 48m  - 28h 10m & 3.5h  & 180s &  60 &  \\
    \hline     
    \end{tabular}
\end{table}

\begin{table}[htbp]
\tbl{The measured equivalent widths and observation date.
(This table is available in its entirety in machine-readable form.)
\\
}{%
\begin{tabular}{@{}lc@{\qquad}lc@{}}  
\hline\noalign{\vskip3pt} 
\multicolumn{1}{c}{Obs. date(JD)} & E.W.({\AA}) & \multicolumn{1}{c}{Obs. date(JD)} & E.W.({\AA}) \\  [2pt] 
\hline\noalign{\vskip3pt} 
July 31\\
2457235.19861 & 3.22    \\
\hline
August 1\\
2457236.13819 & 2.91    &
2457236.14167 & 3.12    \\
2457236.14514 & 3.03    &
2457236.14931 & 2.90    \\
2457236.15278 & 3.09    &
2457236.15764 & 3.02    \\
2457236.16111 & 2.90    &
2457236.16528 & 2.76    \\
2457236.16875 & 2.99    &
2457236.17292 & 2.96    \\
\hline
August 15\\
2457250.19514 &  7.96	 &
2457250.19861 &  7.90	 \\
2457250.20208 &  8.10	 &
2457250.20625 &  8.26	 \\
2457250.20972 &  8.25	 &
2457250.21319 &  8.42	 \\
2457250.21667 &  9.04	 &
2457250.22014 &  10.17   \\
2457250.22431 &  10.36   &
2457250.22802 &  10.47   \\
2457250.23164 &  10.43   &
2457250.23527 &  9.76	 \\
2457250.23888 &  9.87	 &
2457250.24250 &  9.47	 \\
2457250.24625 &  9.66	 &
2457250.25491 &  9.56	 \\
2457250.25852 &  9.66	 &
2457250.26214 &  9.29	 \\
2457250.26575 &  9.25	 &
2457250.26938 &  9.14	 \\
2457250.27300 &  8.70	 &
2457250.27662 &  8.97	 \\
2457250.28024 &  9.10	 &
2457250.28385 &  8.78	 \\
2457250.28748 &  8.31	 &
2457250.29113 &  8.78	 \\
2457250.29476 &  8.41	 &
2457250.29837 &  8.82	 \\
2457250.30204 &  8.63	 &
2457250.30565 &  9.06	 \\ 
\hline
August 26\\
2457261.05948	&    3.95 &
2457261.06171	&    3.86 \\
2457261.06395	&    4.03 &
2457261.06618	&    3.82 \\
2457261.06841	&    3.98 &
2457261.07064	&    4.04 \\
2457261.07287	&    4.00 &
2457261.07509	&    4.05 \\
2457261.07733	&    3.90 &
2457261.07955	&    4.01 \\
2457261.08186	&    4.00 &
2457261.08409	&    4.31 \\
2457261.08632	&    4.17 &
2457261.08855	&    4.29 \\
2457261.09078	&    4.26 &
2457261.09331	&    4.35 \\
2457261.09559	&    4.67 &
2457261.09791	&    4.94 \\
2457261.10013	&    4.68 &
2457261.10236	&    4.83 \\
2457261.10466	&    4.80 &
2457261.10689	&    4.85 \\
2457261.10912	&    4.89 &
2457261.11134	&    4.66 \\
2457261.11358	&    4.51 &
2457261.11595	&    4.68 \\
2457261.11818	&    4.24 &
2457261.12041	&    4.58 \\
2457261.12264	&    4.52 &
2457261.12487	&    4.50 \\
2457261.12726	&    4.22 &
2457261.12948	&    4.37 \\
2457261.13171	&    4.11 &
2457261.13395	&    3.87 \\
2457261.13617	&    3.94 &
2457261.13845	&    2.82 \\
2457261.14068	&    3.74 &
2457261.14292	&    4.34 \\
2457261.14514	&    4.08 &
2457261.14737	&    4.17 \\
2457261.14972	&    4.27 &
2457261.15194	&    4.22 \\
2457261.15418	&    4.33 &
2457261.15640	&    4.43 \\
2457261.15863	&    4.18 &
2457261.16086	&    4.48 \\
2457261.16309	&    4.97 &
2457261.16531	&    5.09 \\
2457261.16755	&    5.08 &
2457261.16978	&    4.88 \\
2457261.17205	&    4.75 &
2457261.17427	&    4.57 \\
2457261.17650	&    4.56 &
2457261.17873	&    4.52 \\
2457261.18096	&    4.63 &
2457261.18328	&    4.42 \\
2457261.18550	&    4.63 &
2457261.18773	&    4.45 \\
2457261.18995	&    4.53 &
2457261.19219	&    4.42 \\
2457261.19456	&    4.64 &
2457261.19679	&    4.39 \\
2457261.19902	&    4.32 &
2457261.20125	&    4.13 \\
2457261.20347	&    4.09 &
2457261.20571	&    4.07 \\
2457261.20794	&    4.42 &
2457261.21016	&    4.36 \\
2457261.21240	&    4.38 &
2457261.21463	&    4.10 \\
2457261.21705	&    4.57 &
2457261.21928	&    4.27 \\
2457261.22150	&    4.36 &
2457261.22374	&    4.10 \\
2457261.22596	&    4.17 &
2457261.22819	&    4.17 \\
2457261.23042	&    4.35 &
2457261.23265	&    4.34 \\
2457261.23488	&    4.56 &
2457261.23711	&    5.24 \\
2457261.24491   &    4.88 &
2457261.24713	&    4.76 \\
2457261.24935	&    4.76 &
2457261.25159	&    4.80 \\
2457261.25382	&    4.71 &
2457261.25605	&    4.63 \\
2457261.25829	&    4.78 &
2457261.26051	&    4.87 \\
2457261.26274	&    4.97 &
2457261.26497	&    4.82 \\
2457261.26725	&    4.79 &
2457261.26948	&    4.65 \\
2457261.27170	&    4.82 &
2457261.27394	&    4.74 \\
2457261.27616	&    4.57 &
2457261.27839	&    4.61 \\
2457261.28061	&    4.35 &
2457261.28285	&    4.46 \\
2457261.28508	&    4.59 &
2457261.28730	&    4.35 \\
\hline                     
August 27\\
2457262.15839 	&    4.55 &
2457262.16063 	&    4.29 \\
2457262.16286 	&    4.38 &
2457262.16508 	&    4.48 \\
2457262.16731 	&    4.35 &
2457262.16958 	&    4.24 \\
2457262.17182 	&    4.00 &
2457262.17405 	&    4.28 \\
2457262.17627 	&    4.40 &
2457262.17851 	&    4.50 \\
2457262.18083 	&    4.24 &
2457262.18307 	&    4.24 \\
2457262.18529 	&    4.33 &
2457262.18752 	&    4.54 \\
2457262.18976 	&    4.54 &
2457262.19755 	&    4.68 \\
2457262.19978 	&    4.58 &
2457262.20200 	&    4.56 \\
2457262.20424 	&    4.79 &
2457262.20646 	&    4.65 \\
2457262.20878 	&    4.69 &
2457262.21101 	&    4.31 \\
2457262.21324 	&    4.24 &
2457262.21546 	&    4.27 \\
2457262.21767 	&    4.20 &
2457262.21993 	&    4.18 \\
2457262.22215 	&    4.34 &
2457262.22439 	&    4.33 \\
2457262.22662 	&    4.15 &
2457262.22884 	&    4.14 \\
2457262.23120 	&    4.06 &
2457262.23343 	&    4.41 \\
2457262.23566 	&    4.00 &
2457262.23788 	&    4.18 \\
2457262.24012 	&    4.32 &
2457262.24234 	&    4.11 \\
2457262.24457 	&    4.10 &
2457262.24679 	&    4.07 \\
2457262.24903 	&    4.08 &
2457262.25126 	&    4.23 \\
2457262.25365 	&    4.06 &
2457262.25587 	&    4.13 \\
2457262.25810 	&    3.97 &
2457262.26034 	&    4.16 \\
2457262.26256 	&    4.21 &
2457262.26493 	&    4.09 \\
2457262.26716 	&    3.93 &
2457262.26939 	&    3.88 \\
2457262.27162 	&    3.78 &
2457262.27385 	&    3.88 \\
2457262.27620 	&    3.82 &
2457262.27843 	&    3.66 \\
2457262.28066 	&    3.58 &
2457262.28289 	&    3.44 \\
2457262.28512 	&    3.69 &
2457262.28762 	&    3.64 \\
2457262.28999 	&    3.94 &
2457262.29222 	&    3.45 \\
2457262.29444 	&    3.62 &
2457262.29682 	&    3.61 \\ 
[2pt]
\hline\noalign{\vskip3pt} 
\end{tabular}}
\end{table}

\end{document}